\documentclass[11pt, svgnames, twocolumn]{article}

\PassOptionsToPackage{hyphens}{url}
\PassOptionsToPackage{colorlinks=true,
  linkcolor=blueviolet, citecolor=crimson,
  urlcolor=coral, filecolor=crimson}{hyperref}

\usepackage[margin=0.85in, top=1.1in, bottom=1in]{geometry}

\usepackage[T1]{fontenc}
\usepackage{lmodern}
\usepackage{microtype}

\usepackage{wrapfig}
\usepackage{multirow}
\usepackage{orcidlink}
\usepackage{listings}
\usepackage{booktabs}
\usepackage{subcaption}
\usepackage{tcolorbox}
\usepackage{graphicx}
\usepackage{float}
\usepackage{etoolbox}
\usepackage{xurl}
\usepackage{algorithm}
\usepackage{algorithmic}
\usepackage{rotating}
\usepackage{array}
\usepackage{colortbl}
\usepackage{cleveref}
\usepackage[utf8]{inputenc}
\usepackage{textcomp}
\usepackage{amssymb}
\usepackage{wasysym}
\usepackage{ragged2e}
\usepackage{verbatim}  
\usepackage[round]{natbib}
\usepackage{hyperref}
\usepackage{fancyhdr}
\usepackage{titlesec}

\newcommand{\sagedoiurl}{https://journals.sagepub.com/doi/10.1177/20539517261467361}
\newcommand{\acceptancenotice}{\small\textit{This article has been accepted for publication by
    \textit{Big Data \& Society} (SAGE). Please cite this version: \url{\sagedoiurl}}}

\fancypagestyle{plain}{%
  \fancyhf{}
  \fancyhead[C]{\acceptancenotice}
  \fancyfoot[C]{\small\thepage}
  
}

\definecolor{crimson}{HTML}{DC143C}
\definecolor{steelblue}{HTML}{4682B4}
\definecolor{rosybrown}{rgb}{0.737,0.561,0.561}
\definecolor{silver}{HTML}{C0C0C0}
\definecolor{maroon}{HTML}{800000}
\definecolor{coral}{HTML}{FF7F50}
\definecolor{blueviolet}{HTML}{8A2BE2}
\definecolor{slategray}{HTML}{708090}

\newenvironment{funding}{%
  \section*{Funding}\addcontentsline{toc}{section}{Funding}\small}{}

\newenvironment{acks}{%
  \section*{Acknowledgments}\addcontentsline{toc}{section}{Acknowledgments}\small}{}

\newcommand{\declaration}[2]{%
  \vspace{6pt}%
  {\noindent\textbf{\textsf{#1}}}\\[1pt]%
  {\small #2}%
  \vspace{4pt}%
}

\providecommand{\keywords}[1]{%
  \vspace{0.8em}
  \noindent\textbf{Keywords:} #1\par}

\titleformat{\section}{\normalfont\sffamily\large\bfseries\raggedright}{\thesection}{0.5em}{}
\titleformat{\subsection}{\normalfont\sffamily\normalsize\bfseries\raggedright}{\thesubsection}{0.5em}{}
\titleformat{\subsubsection}{\normalfont\sffamily\normalsize\itshape\raggedright}{\thesubsubsection}{0.5em}{}

\begin{document}

\twocolumn[{%
\begin{center}
  \vspace{-0.5em}
  {\sffamily\LARGE\bfseries Perpetuating Misogyny with Generative AI:\\[0.4em]
    How Model Personalization Normalizes Gendered Harm}

  \vspace{1.2em}\vspace{2mm}
  \small
  \begin{tabular}{%
    >{\centering\arraybackslash}p{0.44\textwidth}
    >{\centering\arraybackslash}p{0.44\textwidth}}
    \textbf{Laura Wagner}~\orcidlink{0009-0003-4244-6342} &
    \textbf{Eva Cetinic}~\orcidlink{0000-0002-5330-1259} \\[0.2em]
    University of Zurich & University of Zurich \\
    Zurich, Switzerland & Zurich, Switzerland \\
    \texttt{laura.wagner@khist.uzh.ch} & \texttt{eva.cetinic@uzh.ch} \\
  \end{tabular}

  \vspace{1.4em}
\end{center}

\begin{abstract}
Open-source text-to-image (TTI) pipelines have become dominant in the
landscape of AI-generated visual content, driven by technological advances
that enable users to personalize models through adapters tailored to specific
tasks. While personalization methods such as LoRA (Low-Rank Adaptation) offer
unprecedented creative opportunities, they also facilitate harmful practices,
including the generation of non-consensual deepfakes and the amplification of
misogynistic or hypersexualized content. This study presents an exploratory
sociotechnical analysis of CivitAI, the most active platform for sharing and
developing open-source TTI models. Drawing on a dataset of more than 40
million user-generated images and over 230,000 models, we find a
disproportionate rise in not-safe-for-work (NSFW) content and a significant
number of models intended to mimic real individuals. We also observe a strong
influence of internet subcultures on the tools and practices shaping model
personalizations and resulting visual media. In response to these findings, we
contextualize the emergence of exploitative visual media through feminist and
constructivist perspectives on technology, emphasizing how design choices and
community dynamics shape platform outcomes. Building on this analysis, we
propose interventions aimed at mitigating downstream harm, including improved
content moderation, rethinking tool design, and establishing clearer platform
policies to promote accountability and consent.

\medskip
\textcolor{red!60!black}{\textbf{Content warning:} This paper contains
  references to pornography, gender-based digital violence and verbal vulgarity
  that might be disturbing, distressing, and/or offensive.}
\end{abstract}


\vspace{1.5em}
}]

\setcounter{footnote}{0}

\section{Introduction}

The evolution of open-source text-to-image (TTI) technology has shifted from
being a niche area for early adopters to becoming a vast ecosystem driven by
increasing investment and commercialization. The technology continues to advance
rapidly, with recent innovations allowing users to \textit{personalize} models
through model adapters, tailoring them for specific tasks and creative pursuits
such as mimicking a specific art style or reproducing the physiognomy of a
particular individual.

While model personalizations open up unprecedented opportunities for artists and
researchers, they also raise significant ethical and legal concerns, particularly
the potential for misuse such as non-consensual deepfakes and disinformation.
Platforms like~\citet{civitai}, a pioneering and established hub for developing
and sharing open-source TTI models, have played a central role in the expansion
of AI-generated content and specifically TTI models. What sets model-sharing
hubs apart from other platforms is their function: the content shared on these
hubs is not an end product but a tool for generating limitless derivative
content.

In our study we focus on the analysis of CivitAI, a community-driven platform
for sharing, customizing and discovering personalized models. CivitAI differs
from other model-sharing platforms by focusing specifically on creative image
generation, with built-in community features that actively promote model
personalization and derivative visual work, rather than serving as a
general-purpose repository.

The models shared on this platform are usually trained on smaller, more
specialized datasets than general-purpose models, and enable users to create
vast amounts of content that mimics the characteristics of that training data,
which can then be widely shared across social media and other online platforms.
This dynamic harbors a high potential for the rapid propagation of generated
content, the impact of which can extend far beyond the communities creating
these models. Given that such models can produce entirely fabricated,
photorealistic, and convincing imagery within seconds, this capacity is
particularly concerning, as it enables and amplifies forms of
technology-facilitated gender-based
violence~(TFGBV)~\citep{gender-based-digital-sexual-violence-un-women}.

We aim to highlight how CivitAI, as a representative example of
community-oriented model-sharing platforms, may contribute to these concerning
developments by fostering environments that, through certain design choices and
community norms, can become more tolerant of abusive practices while
simultaneously growing less welcoming for some user groups. Drawing on a
definition of misogyny as a set of hostile mechanisms designed to reinforce
patriarchal power structures~\citep{manne2017downgirl}, our objective is to
investigate how these dynamics manifest themselves on open-source model-sharing
platforms, as exemplified by CivitAI. Specifically, we document patterns of
over-sexualization, objectification, and erosion of consent in TTI models and
imagery, with women emerging as a primary target of these
dynamics.\footnote{We use the terms \textit{women} and \textit{girls} to denote
individuals who are perceived or depicted as female, because the harms examined
here are disproportionately directed at such bodies. We recognize gender
diversity and use this terminology for analytical specificity rather than to
exclude other identities.}

\begin{figure*}[ht]
  \centering
  \includegraphics[width=\textwidth]{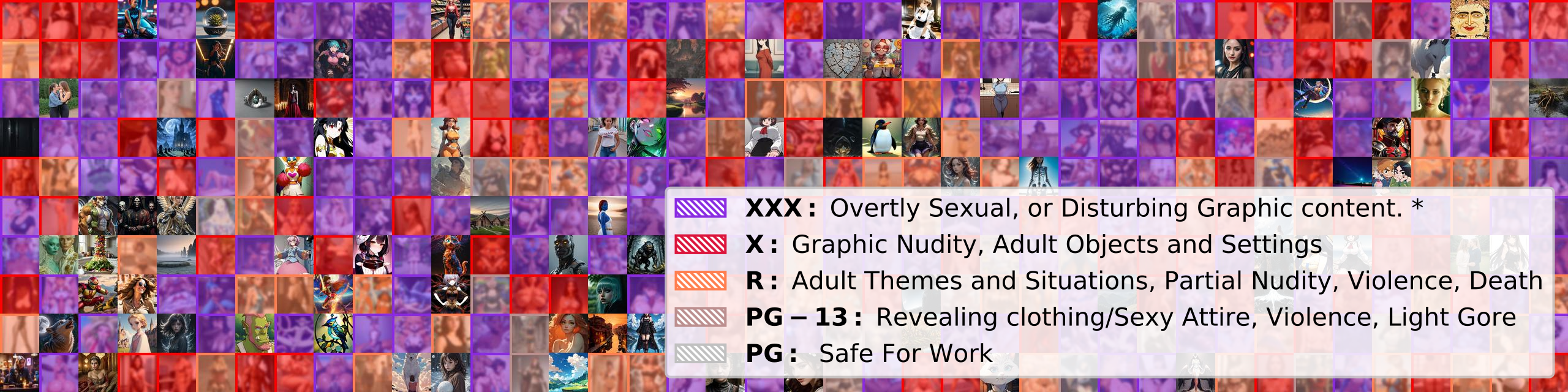}
  \caption{Random sample of images from CivitAI (March 2024) with different
    NSFW levels as assigned by the platform.}
  \label{fig:fig_1_image-grid}
\end{figure*}

Our study examines the underlying mechanisms contributing to the escalation of
NSFW content and the perpetuation of misogynistic tropes within CivitAI. The
prevalence of NSFW content is illustrated in Figure~\ref{fig:fig_1_image-grid},
which shows a uniformly random sample of images from our initial data collection
in March 2024, when we first initiated this study. These images are categorized
by the platform moderation system according to different levels of NSFW and
reflect the distribution of content that initially motivated our analysis. We
perform a detailed analysis of user-generated content, model adapters, and
textual training data to uncover how community-driven, tool-driven, and
model-infused biases may have intensified within the platform's ecosystem over
two years from January 2023 to December 2024. Our analysis reveals patterns
consistent with a self-reinforcing dynamic in which certain types of content
become increasingly normalized. We discuss how these patterns align with
theoretical frameworks suggesting such normalization may foster environments more
permissive of harmful and exploitative material.

Besides documenting problematic aspects of open-source TTI systems, we discuss
potential areas for intervention, such as improving tools used for training
models and model adapters, vetting model metadata, and developing and enforcing
ethical frameworks. Without such interventions, the persistence of problematic
patterns may undermine the legitimacy and long-term sustainability of open-source
innovation, potentially inviting more restrictive external regulation or eroding
public trust in these technologies. By identifying these patterns, our study
aims to contribute to a more inclusive and responsible open-source TTI ecosystem
and to support more equitable representation in AI-generated content.

\section{Investigating Model Personalization and Model-Sharing
  Platforms: A Case Study of CivitAI}

In early 2021, OpenAI introduced the text-to-image system DALL·E
\citep{_dalle_1_nodate} and publicly released its CLIP
component~\citep{radford_clip_paper_2021}, a model trained on large-scale
image--text pairs that enabled semantic alignment between language and images
and became foundational for later text-to-image generation systems. The
open-source community aimed to recreate similar text-to-image functionality
through shared experimentation, with early pipelines circulating via social
media and Discord channels and developing within collaborative communities
focused primarily on creative and artistic applications~\citep{oppenlaender_creativity_2022}.

The release of Stable Diffusion in 2022~\citep{Rombach_2022_CVPR} marked a
milestone as the first open-source TTI foundation model. Its popularity
indicated a shift from community-driven innovation toward investor-funded
commercial activity. At the same time, increased scrutiny emerged over
copyright violations, attribution issues, and social biases stemming from
training on uncurated web-scraped data. Systematic research efforts emerged to
quantify and evaluate gender, racial, and cultural biases in generated images.
In particular, several studies have documented gender biases in occupational
representation that mirror and reinforce real-world
inequalities~\citep{wu_revealing_2025,sim_biased_2025,WuNG24,0001B0WYQPL24}.
Beyond occupational roles, these models have been shown to perpetuate broader
patterns of stereotyping across multiple demographic
categories~\citep{NaikN23,schneider_investigating_2025,sufian_t2ibias_2025}.
Researchers have also identified racial biases in the representation of skin
tones and ethnic
features~\citep{GhoshC23,aldahoul_ai-generated_2025}, as well as cultural
biases that privilege Western perspectives and
aesthetics~\citep{KannenAAPPDBD24,ZhangLYGWYL24,liu_scoft_2024,shi2025culturefadesrevealingcultural}.

Many of these issues emerge from limitations in foundational datasets, which not
only reflect biases but have also been shown to contain various forms of
problematic content. For example, the LAION datasets, foundational to many TTI
models despite being ``not recommended'' for commercial
use~\citep{noauthor_models_all_the_way}, initially contained harmful
content~\citep{birhane_multimodal_2021,birhane_into_2023}, including child
sexual abuse material~\citep{thiel_csam_stanford_2023}. In response, LAION
temporarily removed, cleaned, and re-released the datasets, highlighting that
foundational dataset curation is an ongoing process.

While such interventions address foundation model training data, model
personalizations are created through fine-tuning with small, user-curated
datasets which are mostly undisclosed. As a result, improvements to foundational
datasets have limited impact on the potential harms emerging within
personalization workflows. Recent developments in model personalization
techniques such as LoRA (Low-Rank
Adaptation)~\citep{DBLP:journals/corr/abs-2106-0968-lora}, and textual
inversion~\citep{gal2023imageworthword}, offer refined control over generated
content without retraining entire models. Applications with graphical user
interfaces, such as Stable-Diffusion-WebUI~\citep{automatic1111_stable_2022}
and ComfyUI~\citep{comfyanonymous_comfyanonymouscomfyui_2024}, lowered the
barrier to entry, allowing users to personalize models easily and without
advanced technical skills. However, model personalization introduced new risks,
particularly through the tailored generation of deepfake imagery, exacerbating
digital sexual violence disproportionately affecting women~\citep{viola2023designed}.

Despite the growing prominence of model personalization techniques and sharing
platforms, research into their implications remains limited. Early investigative
journalism documented the prevalence of NSFW content and sexually exploitative
AI-generated imagery, including non-consensual
deepfakes~\citep{maiberg_inside_2023,maiberg_a16z_2023}. Subsequent studies
confirmed the widespread presence of abusive generative practices on CivitAI,
including thousands of non-consensual deepfake
models~\citep{wei2024,hawkins_deepfakes_2025}. Research on user prompting
patterns on CivitAI highlights a high prevalence of NSFW
material~\citep{palmini2024civiverse}. Platform governance research highlights
moderation failures, including weak responses to artists' consent
concerns~\citep{Gorwa2024} and shortcomings in open-source licensing
frameworks~\citep{CuiA24}.

Our exploratory analysis of CivitAI aims to quantify and contextualize these
systemic challenges as they manifest on this platform. Understanding the layered
structure through which content is produced and shaped, including generative
models and the surrounding sociotechnical ecosystem, offers valuable insight
into this case and supports a more holistic perspective on model-sharing
platforms.

\begin{figure*}[t]
  \centering
  \includegraphics[width=0.88\textwidth]{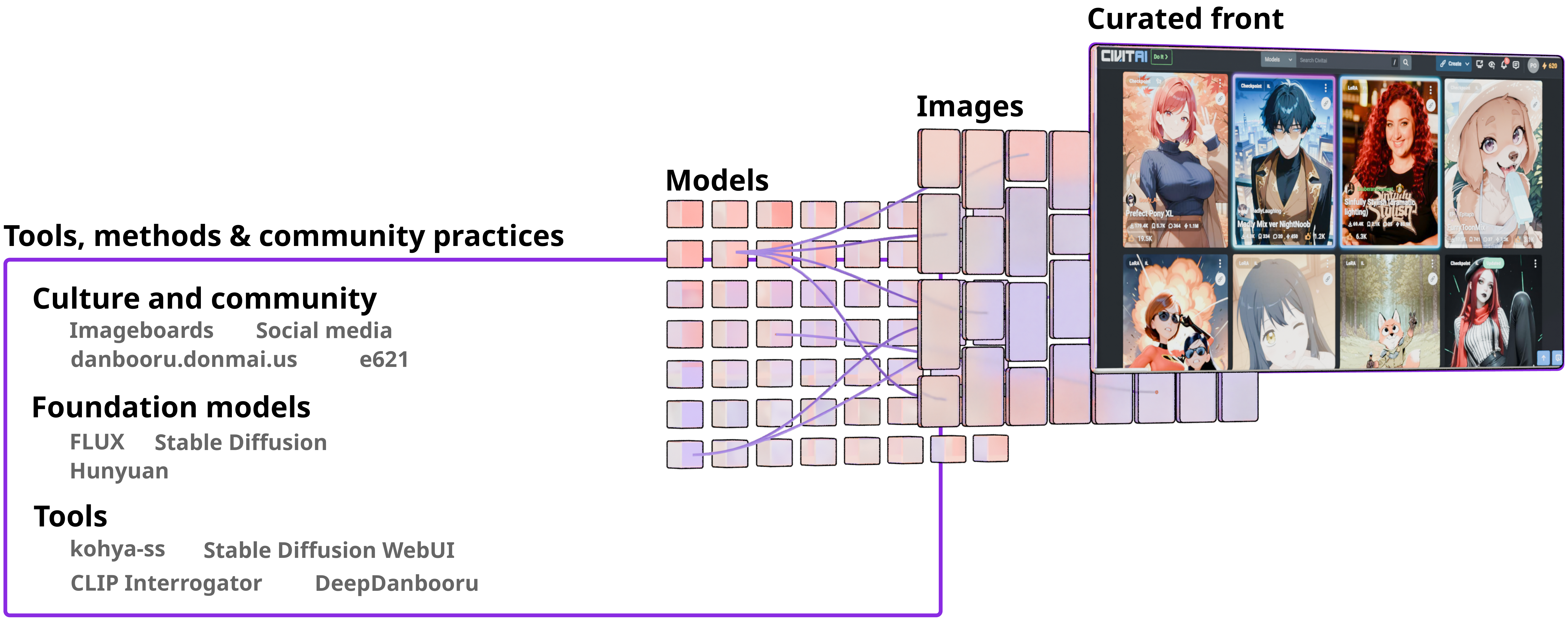}
  \caption{Schematic overview of the CivitAI platform, showing layered
    relationships between the curated front-end images, the underlying
    generative models, and the broader sociotechnical ecosystem. This includes
    tools, data sources, and cultural influences that shape model development.}
  \label{fig:fig_2_layers}
\end{figure*}

Figure~\ref{fig:fig_2_layers} provides a schematic overview of the CivitAI
platform, illustrating the layers of content generation and circulation. We
focused our analysis on CivitAI as it represents the most established
model-sharing platform exclusively dedicated to TTI personalizations,
distinguished by its rapid growth and integration of social community-building
features.

\subsection{Data Collection and Categorization}
\label{sec:3.1_data_collection}

To illustrate the evolving dynamics of this model-sharing platform, we
collected and analyzed image metadata, model metadata, and training process
metadata from content published on CivitAI over a 2-year period (late November
2022 to December 2024), using the CivitAI REST
API~\citep{civitai_rest_api_github}. The dataset covers 40,630,560
images~\citet{wagner2025dataset40m} and 231,252
models~\citet{wagner2025datasetmodels} as of 19\textsuperscript{th} January
2025. The model metadata includes 9,948 standalone models, commonly referred to
as ``checkpoints'', capable of generating images independently, and 221,304
adapters (e.g., LoRA, Textual Inversion) that require a checkpoint to function.
We analyzed three primary data categories: \textit{image metadata},
\textit{model metadata}, and \textit{training process metadata}.

Image metadata consists of \textit{content-organization metadata} and
\textit{generation metadata}. Content-organization metadata includes identifiers
(CivitAI ID, perceptual hash), creator usernames, timestamps, engagement metrics
(likes) and NSFW categories. CivitAI assigns images to one of five NSFW levels
using a tiered content classification system known as ``browsing levels'', loosely
based on Motion Picture Association film ratings~\citep{motion_picture_ratings_film_nodate},
ranging from level~1 (PG, safe for all ages) to level~8 (X, restricted to
adults~18+), with an additional level~16 (XXX) introduced for the most extreme
content. Generation metadata includes the settings, models, and adapters used to
produce each image (identified by CivitAI IDs or model hashes). For an example
structure, see Figure~\ref{fig:image_metadata} in the Supplementary Material.

Model-related metadata includes unique CivitAI IDs, version-specific IDs,
download counts, timestamps, engagement metrics and a POI (person of interest)
flag indicating deepfake models. Metadata from model creation pipelines,
including base model details and framework-specific hashes, facilitates
identification. A detailed structure is shown in Figure~\ref{fig:model_metadata}
in the Supplementary Material.

Finally, we extracted metadata written during training of a model, which were
used to identify the auto-tagging system responsible for annotating the training
data for model personalizations. This metadata was extracted from the downloaded
model adapter files (in the \texttt{*.safetensors}
format~\citep{by321_by321safetensors_util_2024}). A detailed overview of the
data structure can be found in our Supplementary Material,
Figure~\ref{fig:metadata_safetensors}.

\subsection{Image-Level Analysis: Trends and Representations}

To deepen our understanding of the visual material beyond initial impressions of
the user-facing layer of the CivitAI platform, we examined all images and their
metadata shared between January 2023 and December 2024. We first quantified
temporal trends in the distribution of NSFW classification levels assigned by
the CivitAI platform, as extracted from image metadata. We then analyzed the
images themselves with regard to the demographic characteristics of depicted
subjects, employing a classifier trained to infer age and gender to reveal
broader trends in the generation of human figures.

\subsubsection{Quantifying NSFW Trends}

We examine how the composition of the CivitAI image corpus has changed over
time with respect to associated NSFW levels. A previous study by
\citet{palmini2024civiverse} revealed a disproportionate increase in NSFW
imagery compared to SFW on CivitAI, with the NSFW ratio increasing from 56\%
in October 2023 to 73\% in April 2024. Additionally, \citet{wei2024}
demonstrated that NSFW images receive more social engagement on CivitAI than
SFW images.

\begin{figure*}[t]
  \centering
  \begin{minipage}{0.47\textwidth}
    \centering
    \includegraphics[width=\textwidth]{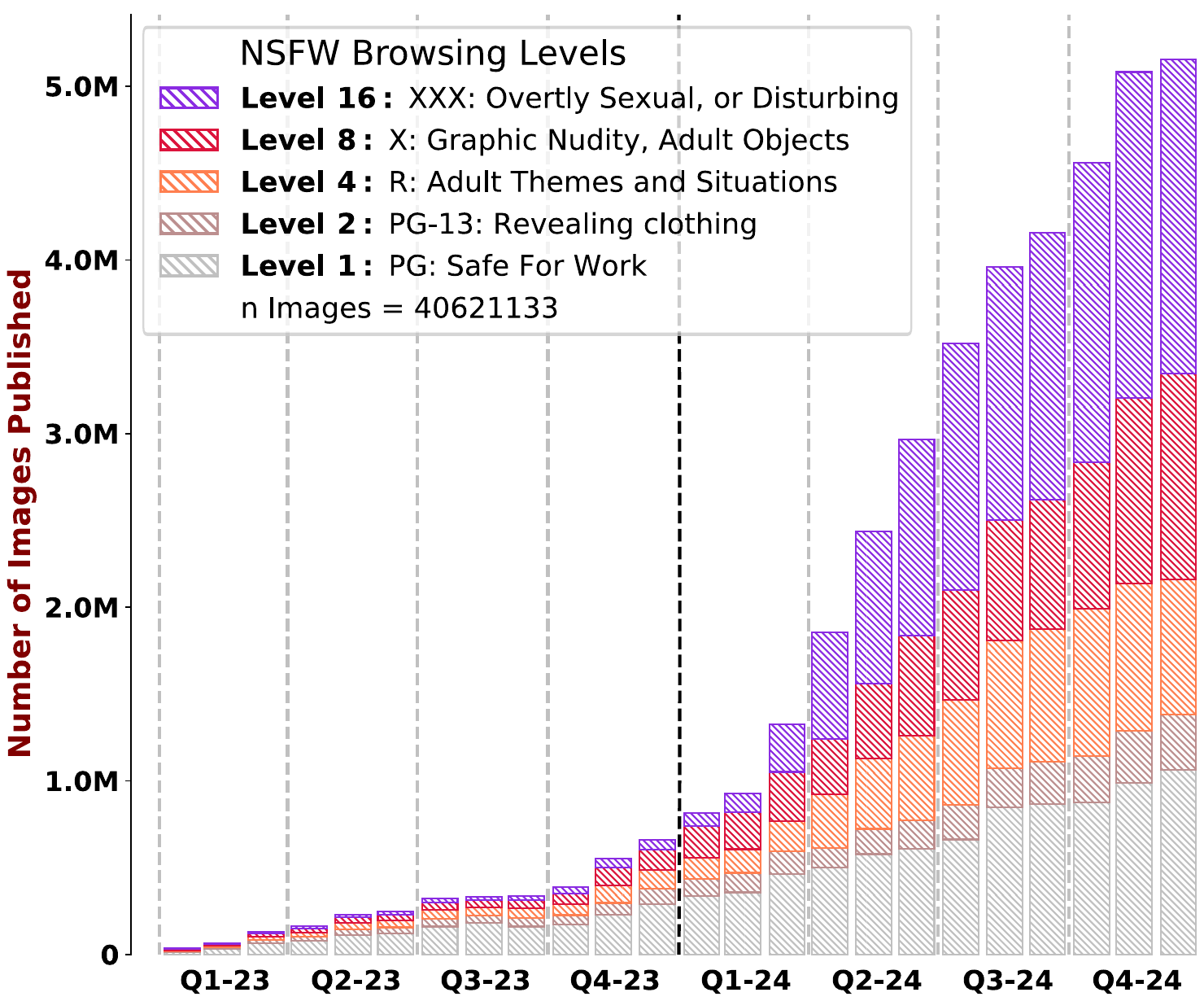}
    \caption{Monthly distribution of generated images categorized by the NSFW
      browsing level assigned by CivitAI.}
    \label{fig:3a_nsfw_ratio_images}
  \end{minipage}\hfill
  \begin{minipage}{0.48\textwidth}
    \centering
    \includegraphics[width=\textwidth]{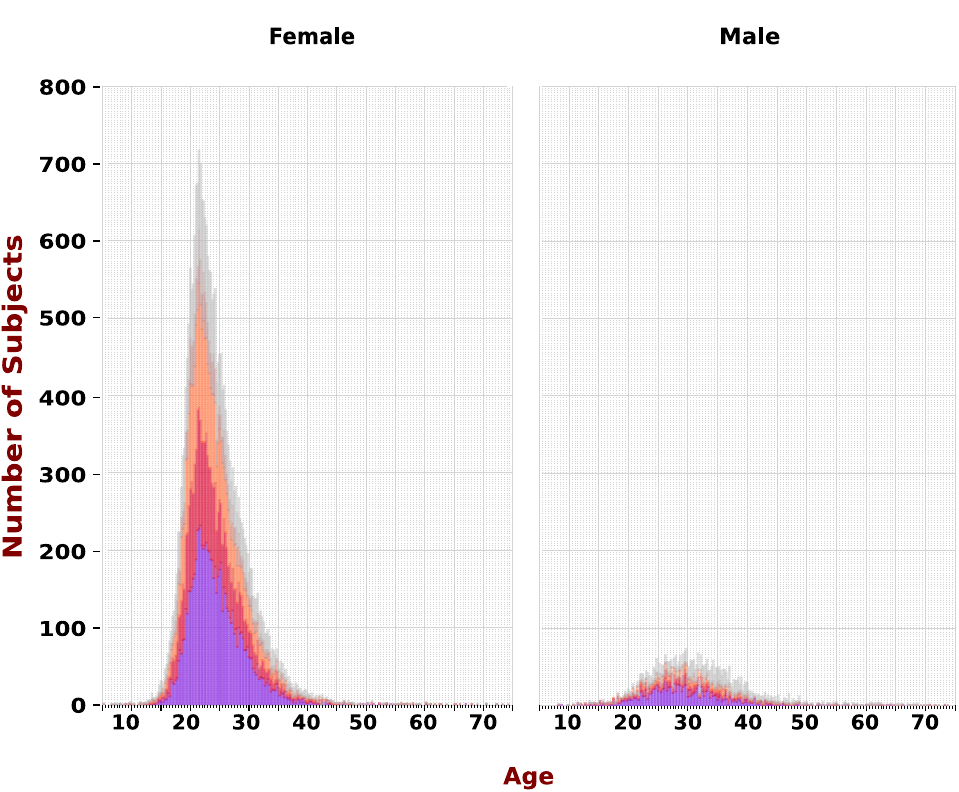}
    \caption{Inferred age and gender predictions using MiVOLO on a 0.1\%
      sample of temporally representative images from the 40 million image
      dataset.}
    \label{fig:3b_mivolo_gender_age}
  \end{minipage}
\end{figure*}

Our current study extends the dataset to include model metadata and covers a
full two-year period, from January 2023 to December 2024. With this extended
dataset, we confirm a continuation of the trend, with the NSFW ratio increasing
from 41\% in January 2023 to 80\% by the end of December 2024. This trend is
visualized in Figure~\ref{fig:3a_nsfw_ratio_images}, which presents a stacked
bar diagram showing the color-coded proportions of over 40 million images
according to their assigned NSFW browsing level, posted to CivitAI between
January 2023 and December 2024. A more detailed account of monthly trends in
image generation and NSFW classification is provided in
Table~\ref{table:civiverse-dateset-table} in the Supplementary Material.

\subsubsection{Demographic Patterns in Generated Images}

To analyze and quantify the demographic distribution of human subjects depicted
in generated images shared via CivitAI, we employed the \textit{MiVOLO}
(Multi-Input VOLO) model introduced by~\citet{kuprashevich_mivolo_2023}. This
is a transformer-based system designed to predict age and gender from images,
even when faces are partially or fully occluded, while excluding images without
human subjects. The results are shown in Figure~\ref{fig:3b_mivolo_gender_age},
which shows the distribution of estimated gender and ages and the corresponding
NSFW browsing levels.

The gender and age estimation was performed on a sampled subset of the original
40-million-image dataset. The subset consists of 40,636 images, representing
approximately 0.1\% of the full dataset. It was created through sampling to
ensure a temporally balanced and distributionally representative sample. Using
the MiVOLO classifier, we identified 29,073 human subjects in the 40,636
images, discarding images with no human subjects or a low confidence level for
gender classification.

The results of gender classification indicate a stark over-representation of
female subjects, with an average female-to-male ratio of 6.24 to 1 across the
dataset. NSFW co-occurrence rates were also notably higher for female subjects,
with 85.44\% of images depicting women having a browsing level other than~1,
compared to male subjects, where the NSFW incidence was 73.66\%. The average
age for female subjects was 23.67 years, with a standard deviation of 5.1,
while male subjects had a higher average age of 31.16 years with a standard
deviation of 9.09 years. This indicates a narrower age range and a strong
emphasis on sexualized portrayals of younger women, in contrast to a broader
age distribution and more varied depictions of male individuals. A more detailed
overview of these results is given in
Table~\ref{appendix:table_mivolo_2023_2024} in the Supplementary Material.

Several limitations should be noted: CivitAI assigns browsing levels using
Amazon Rekognition, unspecified open-source NSFW classifiers, and manual
reviews. NSFW classifiers disproportionately misclassify images of female
subjects as explicit compared to male
subjects~\citep{leu_auditing_2024}. Additionally, MiVOLO's binary gender
classification relies on visually inferred traits, ignoring non-binary and
gender-diverse identities, and may produce uneven error rates, especially for
darker-skinned individuals~\citep{pmlr-v81-buolamwini18a}. Age estimation
models also introduce biases; for instance, digital ageism can distort the
representation of older adults, particularly older
women~\citep{chu22ageism}. While the classifier excludes images without
recognized human subjects and we employed conservative thresholds to reject
borderline cases, AI-generated images often exhibit hybrid characteristics that
blend cartoonish and photorealistic elements. In some cases, images with
stylized or cartoonish features may have been classified as depicting human
subjects, potentially introducing some distortion in the reported demographic
distributions.

\subsection{Model-Level Analysis: Themes and Biases}

While generated images often attract the most attention, it is important to
recognize that platforms like CivitAI primarily function as model-sharing hubs.
Of the 232,164 assets published on CivitAI between November 2022 and January
2025, only 4.28\% were standalone \textit{checkpoint} models, while 94.04\%
were \textit{model adapters}, which require a compatible checkpoint to generate
images. A breakdown of asset types appears in
Figure~\ref{fig:assets-shared-on-civitai} in the Supplementary Material. To
examine the thematic focus of these models, we began with an analysis of
user-generated promotional tags. We then evaluated gender bias reproduction by
running inference tests on the ten most popular checkpoint models. Next, we
examined thirty popular model adapters, assessing their function in augmenting
the visual repertoire of base models. Within this group, we identified a
specific subset, referred to as \textit{deepfake adapters}, which were designed
to replicate particular individuals and made up about 15\% of all adapters.

\begin{figure*}[t]
  \centering
  \includegraphics[width=0.85\textwidth]{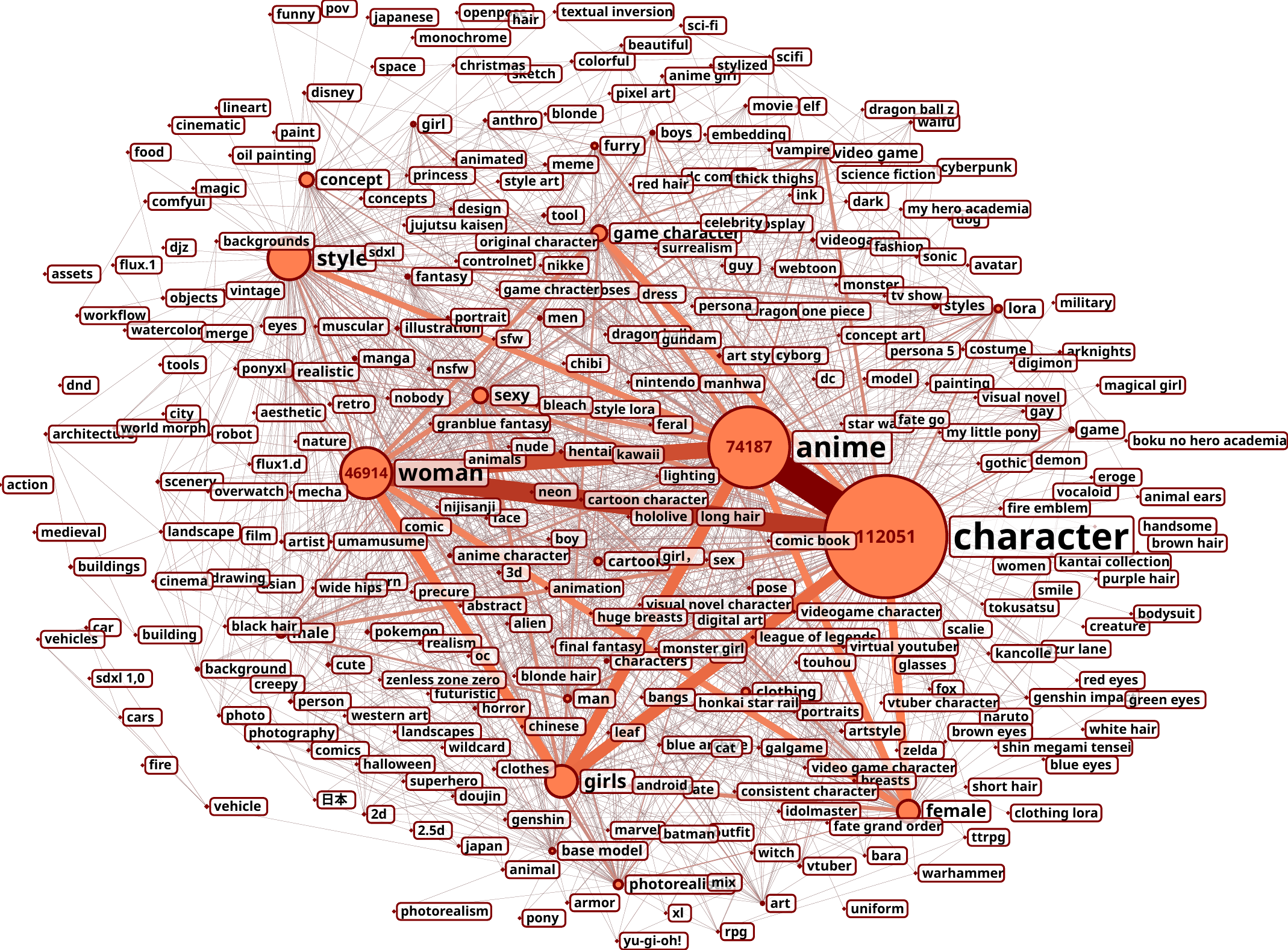}
  \caption{Co-occurrence network of the 300 most frequent promotional tags
    across all assets shared on CivitAI. Node size indicates frequency; edge
    thickness and opacity indicate co-occurrence strength. Interactive version
    available at: \url{https://pm-paper-b68c5e.pages.uzh.ch/figure_5.html}}
  \label{fig:fig_4_tags_co-occurence_network}
\end{figure*}

\subsubsection{Thematic Overview of Models Shared on CivitAI}

To understand dominant thematic orientations, we collected promotional tags
added by model creators on CivitAI. These tags categorize and describe thematic,
aesthetic, and functional attributes of models and adapters during the publishing
process. We analyzed their frequency and co-occurrence across the platform.
Figure~\ref{fig:fig_4_tags_co-occurence_network} presents a co-occurrence
network of the 300 most common tags. Each node represents a tag, with size
indicating frequency. Edges indicate co-occurrence, with thicker, darker lines
representing stronger connections.

The most active cluster revolved around ``character'', occurring in 50.39\% of
models. This is followed by the tag ``anime'', appearing in 32.18\% of models,
indicating that anime aesthetics is a prevalent theme on the platform.
Additionally, tags denoting female gender were highly prevalent. The tag
``woman'' appeared in 29.2\% of models, ``girls'' in 15.7\%, and ``female'' in
11.2\%. While these tags often co-occur in individual models, their high
prevalence indicates a pronounced emphasis on women and girls as depicted
subjects across all models exchanged on the platform. 20.97\% of all models
tagged ``woman'' were also tagged ``sexy'', while this term co-occurred in only
1.34\% of adapters tagged ``man''.

\subsubsection{Bias in Model Checkpoints}

To visualize what embedded discrimination in popular diffusion models means in
terms of visual output, we performed an experiment on the top ten most popular
checkpoints ranked by download count. We prompted each checkpoint model
separately with the single-word prompts \textit{``woman''} and \textit{``man''},
using consistent parameters: CFG scale 6.5, Euler
sampler~\citep{qi2024noisescreatedequallydiffusionnoise}, and 15 inference
steps. To ensure comparability, we initialized the generation process with the
same random noise seed for each model.

\begin{figure*}[t]
  \centering
  \includegraphics[width=0.85\textwidth]{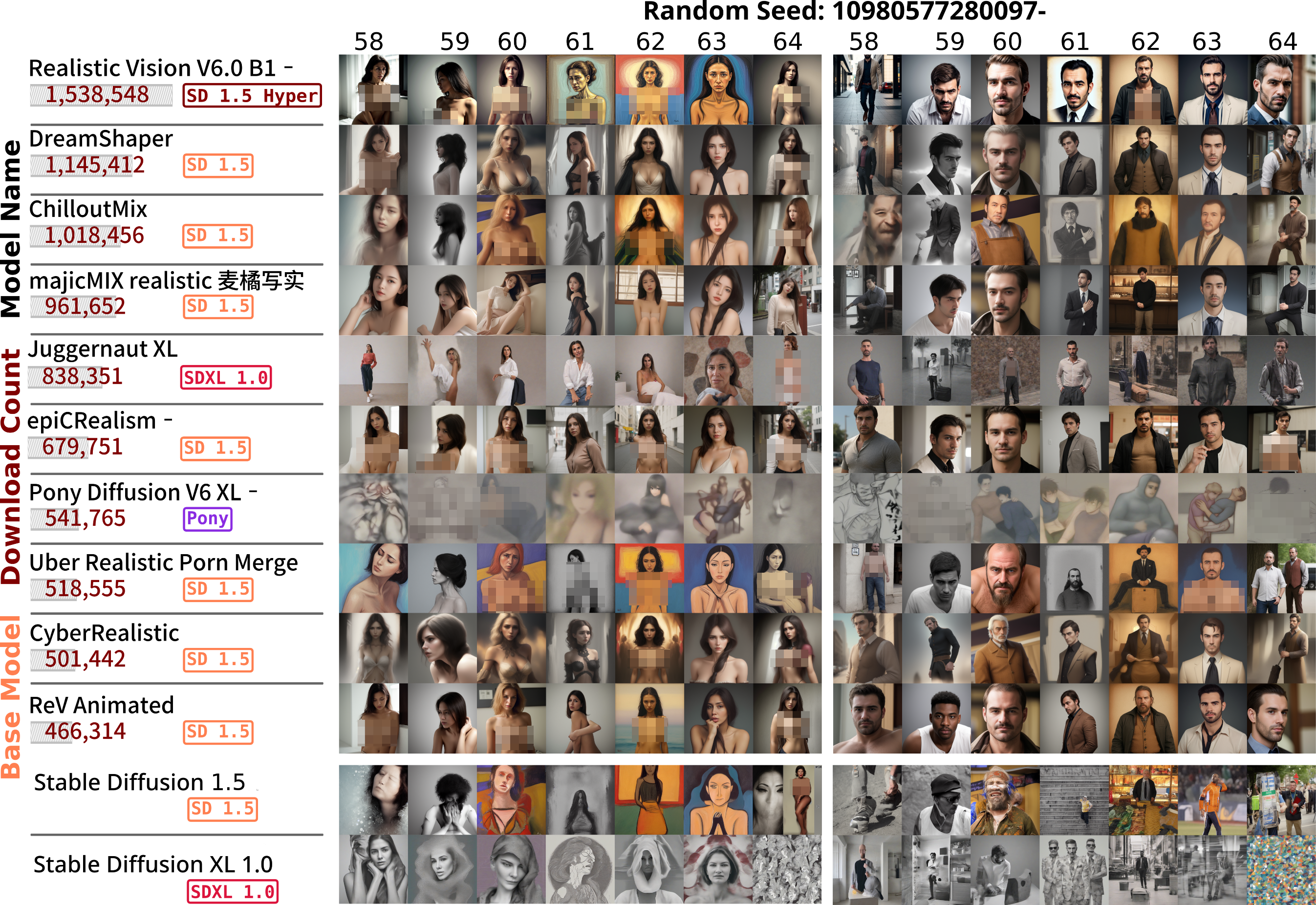}
  \caption{Top 10 most popular standalone models (checkpoints) as of
    19\textsuperscript{th} January 2025, prompted with the neutral terms
    \textit{woman} and \textit{man} using identical parameters. Photorealistic
    and semi-realistic nudity was pixelized.}
  \label{fig:models_inference}
\end{figure*}

The selected checkpoints and their inference results are visualized in
Figure~\ref{fig:models_inference}. For comparison, we also included results
from Stable Diffusion 1.5 and Stable Diffusion XL 1.0, since most of the
selected checkpoints (aside from Pony Diffusion) are derivatives of these
models. The generated outputs across all models revealed a stark difference in
gender representation: women were predominantly depicted nude, while men were
frequently portrayed in professional attire. While the baseline models still
exhibit noticeable gender biases, these biases are even more pronounced in the
model derivatives popular on CivitAI.

\subsubsection{Most Popular Model Adapters}

The core aim of model personalization is to enable users to generate synthetic
imagery with a degree of specificity and control that general-purpose models
cannot achieve through prompting alone. This raises an important question: what
kinds of ``personalized visual worlds'' are being created and shared in practice
most commonly? To explore this, we examine the most popular model adapters on
CivitAI. We identified the 30 most downloaded adapters on
19\textsuperscript{th} January 2025, along with their promotional tags, the
compatible base architecture, and download counts.

\begin{figure*}[t]
  \centering
  \includegraphics[width=0.85\textwidth]{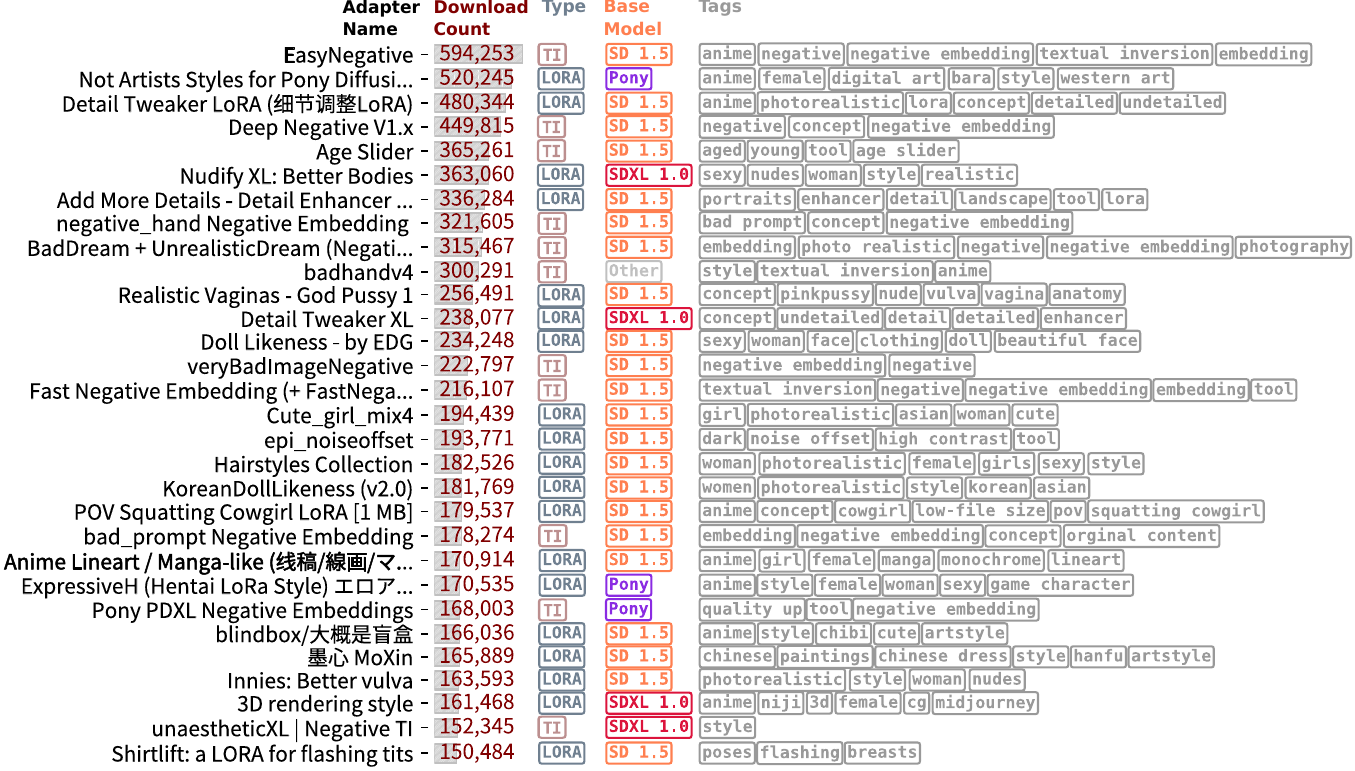}
  \caption{Top 30 most popular model adapters on CivitAI, including the number
    of downloads, adapter type, associated base model, and user-assigned
    promotional tags. Textual Inversion was abbreviated with ``TI''.}
  \label{fig:lora_model_adapters_user_promoted_tags}
\end{figure*}

A brief examination of the promotional tags of the most popular models confirms
the thematic orientation identified in our broader tag analysis, with frequent
references to tags such as ``anime'', ``female'' and ``sexy'', suggesting a
strong emphasis on gendered and sexualized representations.

Indeed, a closer look at the 30 most downloaded adapters reveals a strong
presence of adapters implying pornographic themes. Many of the promotional tags
accompanying these models specifically emphasize female subjects or body parts,
with three models focused solely on the depiction of vaginas. Disconcertingly,
one of the most popular adapters, \textit{Nudify XL: Better Bodies}, directly
references the practice of non-consensual ``nudification'', where photos of
dressed individuals are manipulated to appear naked.

Surprisingly, even models focused on stylistic variation often include tags like
``female''. While CivitAI does not disclose how tags influence visibility, it is
likely that popular tags improve discoverability. This can create a feedback
loop: widely used terms like ``female'', ``girl'', or ``sexy'' not only reflect
user interest but also shape platform norms, as creators adopt them to boost
exposure.

\subsubsection{Deepfake Adapters and Targeted Subjects}

One of the most ethically concerning applications of model personalization is
the creation of model adapters intended to replicate real individuals. To better
understand the scope and impact of this phenomenon, we focused on analyzing the
subset of such adapters, investigating their prevalence, common use cases, and
the demographic characteristics of the individuals they target. According to a
report by~\citet{deeptrace2019state}, 96\% of all deepfake content online was
pornographic, a trend that intensified by 2023 when a report by Home Security
Heroes found that 98\% of deepfake videos were pornographic and 99\% of the
victims were women~\citep{homesecurity2023deepfakes,birrer2024we}.

Although CivitAI prohibits users from publicly sharing explicit content
involving real people, it has no control over how models are used once
downloaded or what is shared on external platforms. This creates a significant
blind spot in the enforcement of ethical guidelines, as deepfake models can
still be deployed in harmful ways beyond the platform's visibility. Given this
regulatory gap and the continuing risk of abuse, we sought to understand who is
being targeted by these models.

To gain more insight into the individuals currently targeted by the practice of
non-consensual deepfakes created from models shared on CivitAI, we annotated
our dataset of deepfake adapters using large language models, similarly to the
approach pioneered by~\citet{wei2024}. This allowed us to identify the most
commonly associated professions and countries linked to those individuals.

\begin{figure*}[t]
  \centering
  \includegraphics[width=\textwidth]{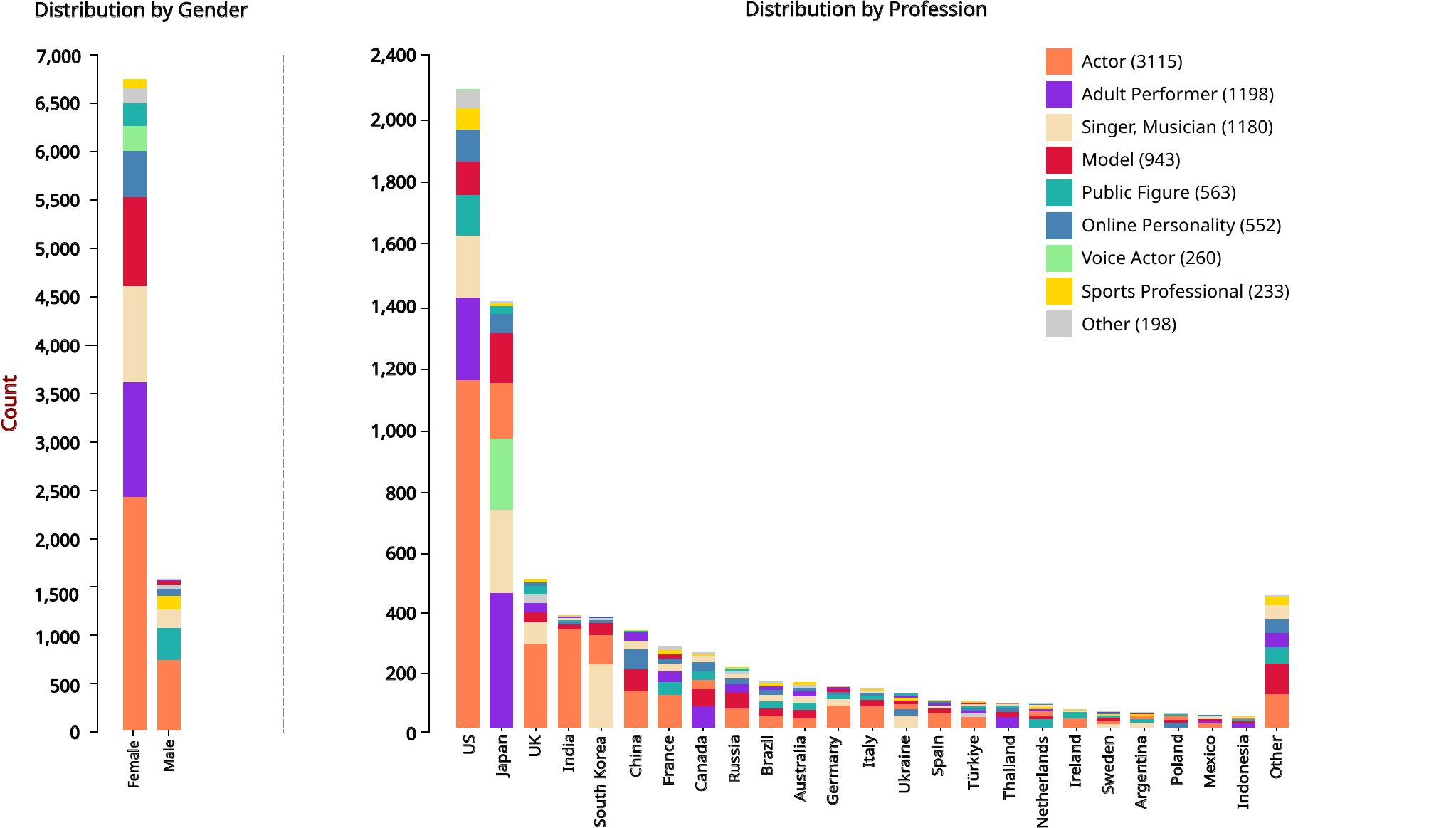}
  \caption{Bar charts depicting the distribution of 8,243 individuals targeted
    by deepfake models available on CivitAI, categorized by LLM-predicted
    profession and gender (left), and associated country (right).}
  \label{fig:Fig_7_professions_genders}
\end{figure*}

In our dataset of models and adapters, we found that 15.27\% (33,804 model
adapters) were tagged ``person of interest'', meaning they were tailored to
replicate identifiable individuals. The individual's name is rarely included in
the model tags and is typically indicated in the model name.

To find more data on the individuals targeted by deepfake models, we
standardized model filenames by converting obfuscated text into regular
characters and removing version numbers, emojis, and model-adapter-related
terms such as ``LoRA'' or ``SD1.5''. We then used the
\verb|en_core_web_trf|~\citep{spacy_en_trf} model from spaCy
3~\citep{spacy3} for named entity recognition (NER) to extract cleaned names
from the processed filenames. Subsequently, we queried three large language
models to annotate each entry:
Mistral-Small-Instruct-2409~\citep{mistralai2024mistralsmall},
Gemma-3-27b-it~\citep{geminiteam2024geminifamilyhighlycapable}, and
Qwen2.5-32B-Instruct~\citep{yang2024qwen2technicalreport}. The prompt was
constructed from the NER-identified name and contextual hints derived from the
model's promotional tags, instructing the LLMs to provide the full name of the
targeted person, the country name they are most associated with, their
associated gender and up to three different professions. To ensure conservative
predictions, the models were instructed to return ``unknown'' whenever uncertain
and were configured with a temperature of 0.1 and top\_p of 1.0 at inference.

We applied a multi-stage filter. First, we kept annotations only when at least
two of the three models agreed on the individual's name and country, allowing
for minor variations such as name order. Second, we kept only entries where all
three models agreed on gender. We removed rows that were uncertain, referred to
fictional characters, or were trained on multiple individuals. Finally, we
merged all models linked to the same person so each individual appeared only
once. The resulting dataset of 8,243 individuals is shown in two bar charts in
Figure~\ref{fig:Fig_7_professions_genders}.

Our results indicate that actors emerged as the most common profession for both
genders. Among women, adult performers were the second most frequent category,
whereas public figures ranked second among men. The majority of targeted
individuals were associated with countries in the Global North, specifically the
United States, the United Kingdom, and Japan. Cultural factors appear to
influence who is targeted: Japan showed a notably high presence of voice actors,
likely reflecting the country's strong anime and gaming industries. Sports
professionals were also frequently identified, a pattern that may be attributed
to the increased media exposure surrounding international events such as the
2024~Olympics.

This analysis offers insight into the cultural and professional context of
deepfake victims but remains constrained by methodological limits. Language
models may misclassify individuals or overlook multiple roles, and the
experiment is inherently limited to recognizing individuals with a public
record. Even with these limitations, the results provide a structured entry
point for examining the social dynamics behind deepfake creation.

The creation of deepfake explicit images is an abuse that highlights the risks
of model personalization. Yet beyond these malicious uses lies a systemic
problem: the spread of misogynistic and ethically troubling content through
model training and dissemination. These issues often arise not from intent, but
from seemingly neutral processes like image-text pairing.

\subsection{System-Level Analysis: Configurations and Training Practices}

Even when personalization workflows do not explicitly aim to sexualize or
objectify individuals, the training mechanisms themselves can introduce and
reinforce such representations.

Training LoRA adapters requires image--caption pairs. Because manually writing
captions is time-consuming, auto-tagging models are widely used. Although many
TTI models exist, only a few captioning models are commonly adopted.

One widely used system is CLIP Interrogator, which generates captions by
comparing image features with a structured vocabulary through the CLIP
model~\citep{radford_clip_paper_2021}. Its vocabulary is organized into
categories such as artists, mediums and art
movements~\citep{pharmapsychotic_pharmapsychoticclip-interrogator_2025}. It
includes a few violence-related terms but no explicit sexual terms.

The other major approach relies on Danbooru-based taggers.
\citet{danbooru} is an anime-focused imageboard with a tag set that contains
many sexually explicit terms and comparatively few labels for domains such as
nature or art history~\citep{britt_danbooru_content_2019}. Several tagging
models have been trained on this dataset, including
DeepDanbooru~\citep{kim_kichangkimdeepdanbooru_2024}. Danbooru-based models are
integrated into popular LoRA training tools such as
kohya-ss~\citep{bmaltais_bmaltaiskohya_ss_2025} and Stable Diffusion WebUI. In
ComfyUI, users can install extensions such as WD-14
Tagger~\citep{smilingwolf_wd-14-tagger_tagscsv_2022}.

Danbooru-based auto-captioning models typically produce short tag sequences such
as \textit{1girl, red dress, white background}, while CLIP Interrogator
generates full-sentence descriptions. These distinct patterns allow us to
determine which model was used. To categorize training keywords and identify
sensitive content, we created a structured taxonomy of Danbooru tags based on
the public tag groups on the site. The resulting hierarchy of 35,559 terms
appears in Figure~\ref{fig:Fig_14_danbooru_taxonomy}.

To examine auto-tagging practices on CivitAI, we analyzed the 40,000 most
downloaded models. For about 11,000 models, we could identify the tagging system
by comparing metadata to the characteristic vocabularies and structures of CLIP
Interrogator and Danbooru-based models. If fewer than five matches were found or
captions were missing, the system was marked as unknown.

Only 14\% of adapters matched CLIP Interrogator patterns, while more than 70\%
matched Danbooru-based vocabulary. Among those models representing real
individuals, 42.6\% included Danbooru-derived tags.

\begin{figure*}[t]
  \centering
  \includegraphics[width=\textwidth]{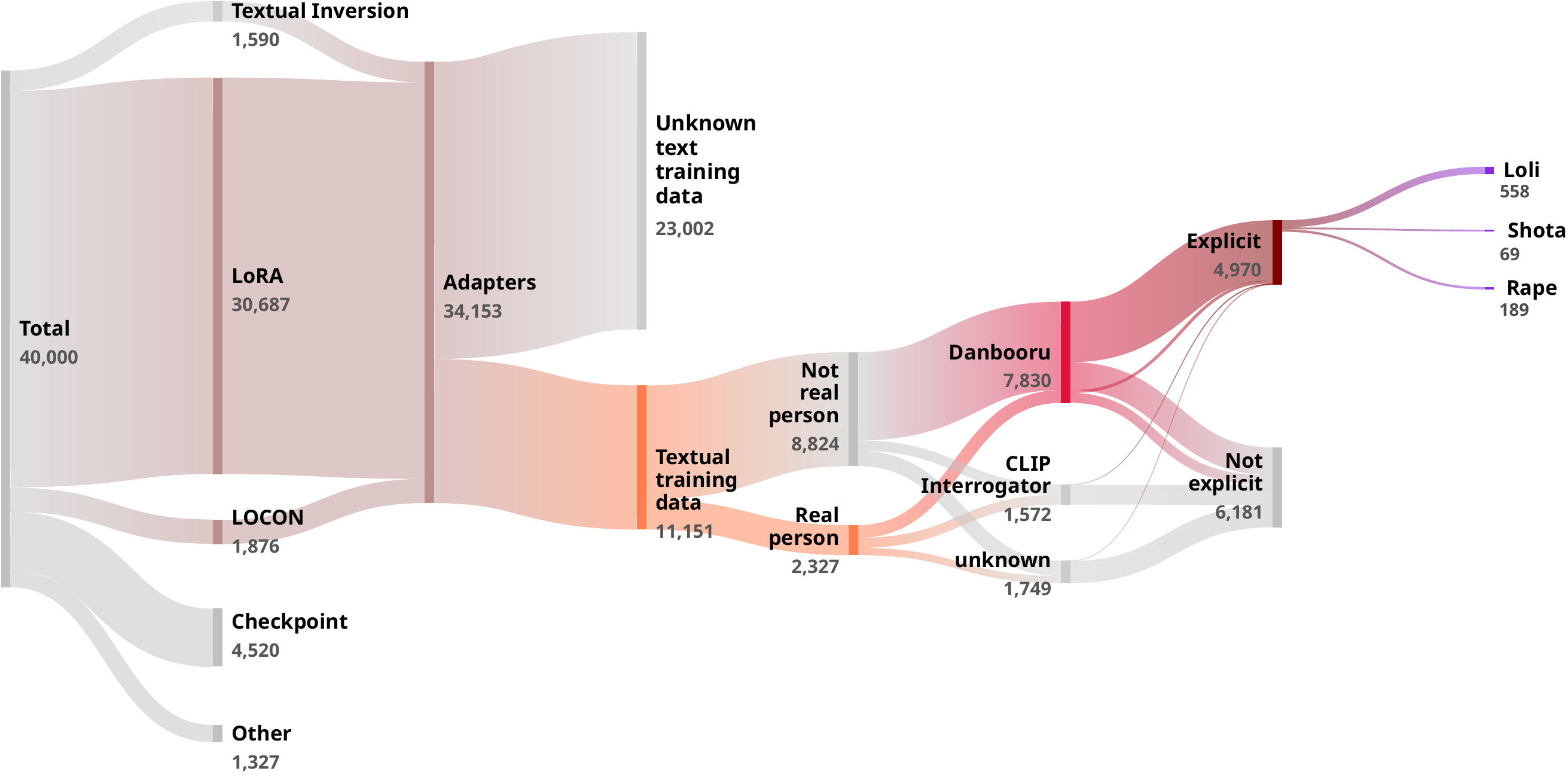}
  \caption{Subdivision of the 40,000 most downloaded CivitAI assets by type,
    presence of extractable training data, and occurrence of problematic tags.}
  \label{fig:sankey_40000}
\end{figure*}

Figure~\ref{fig:sankey_40000} illustrates the overall landscape, showing how
many of the 40,000 most-downloaded models are adapters, how many contain
extractable training text, and within that group, how many include explicit or
abusive tags. Among models with extractable captions, about half contained at
least one tag mapped to the Danbooru ``sex'' category. Roughly 5.6\% included
the keywords ``loli'' and ``shota'', which typically refer to sexualized
depictions of minors, and 2.1\% included ``rape''. These findings indicate that
captioning tools and training configurations have a significant influence on TTI
personalization. They also show how auto-tagging models embed norms from niche
online communities, normalizing sexualized depictions of minors and reinforcing
misogynistic tropes.

\section{Sociotechnical Implications and Pathways for Intervention}

The following discussion is organized around four principal thematic domains
identified through our analysis: content degradation and platform decay driven
by engagement-oriented incentives; the amplification of gendered biases and
misogyny in generative outputs; the proliferation of non-consensual deepfakes
enabled by accessible tools; and potential avenues for intervention through
technical, regulatory, and community-based approaches.

\subsection{Commercialization and Content Degradation}

What began as a promising creative breakthrough in TTI generation and model
personalization has, on platforms like CivitAI, devolved into a pipeline for
the large-scale production of sensational, biased, and abusive content. The
open-source nature of TTI technologies, proclaimed as a democratizing force in
generative AI, has also enabled the propagation of models that perpetuate
hypersexualized imagery and non-consensual deepfakes. This shift reflects a
broader structural trend described by the phenomenon of
\textit{enshittification}, as coined by~\citet{doctorow_platform_2025}, also
known as platform decay. It describes the process by which a platform originally
offering value to its customers grows increasingly exploitative under the
imperative to recoup initial investments.

While Doctorow's framework centers on venture-capital-driven commercialization,
model-sharing platforms like CivitAI present a distinct variant of these
dynamics. The platform actively incentivizes sensational content through its
reward economy: users can credit creators with the platform's virtual currency
``Buzz''~\citep{civitai_buzz_guide} and vote to feature models in online
generators. This gamified system creates financial and reputational incentives
that favor engagement-maximizing content, which combined with insufficient
moderation and network effects, drives trajectories toward harmful content
concentration.

Platform toxicity often extends beyond internal dynamics into broader social
contexts~\citep{trott_operationalising_toxicity_2022}. Social media entities are
explicitly marketed as platforms rather than publishers, a distinction that
suggests they are not bound by the ethical responsibilities typically expected
of traditional publishers~\citep{applebaum2024autocracy}. For model-sharing
platforms, the consequences of this distinction may be especially problematic,
as the output of models and model adapters, by design, can serve as means to
generate and perpetuate content across the internet.

Open-source technologies frequently transition from collaborative,
community-driven origins into profit-oriented ecosystems~\citep{widder_open_2023},
a pattern TTI models now follow. These dynamics have already begun to trigger
external intervention, as illustrated by the recent decision to restrict access
to CivitAI in the UK~\citep{blocked_osa_blocks_civitai_2025}, underscoring how
unchecked platform practices can prompt regulatory responses that impact the
open-source TTI ecosystem. Several indicators suggest a self-reinforcing
feedback loop accelerating these dynamics, including a dramatic increase in NSFW
imagery from 41\% to 80\% in two years, as well as the community's
normalization of deepfakes, misogynistic tropes, and other exploitative content
as documented in this study.

\subsection{Perpetual Misogyny}

According to \citet{manne2017downgirl}, misogyny is not merely the hatred of
women or the ideological framing of women and girls as inferior, which
constitutes sexism. Rather, it is performative: a hostile set of mechanisms and
actions designed to reinforce patriarchal power structures. The
over-sexualization and objectification observed in our analysis exemplify such
mechanisms, as they perpetuate harmful stereotypes and contribute to a culture
that marginalizes women and may deter their participation on the platform.

Drawing on Mulvey's notion of the ``male gaze''~\citep{Mulvey1989} and Berger's
critique of Western art~\citep{berger_ways_2008}, a recent
article~\citep{The_algorithmic_gaze_king} elaborates on how AI-generated imagery
echoes the historical objectification, commodification and hypersexualization of
female bodies. Such visual compartmentalization of the female body aligns with
documented differences in cognitive processing associated with
objectification~\citep{objectification_gervais_pdf_2024}.

Our analysis of the most popular model adapters indicates the persistence of
this trend in the context of generative AI personalization, where women are
frequently depicted through fragmented, sexualized representations that
emphasize specific body parts. The reduction of women to isolated body parts
through objectification extends into fantasies of replacement, where the human
subject is erased and substituted with artificial representations devoid of
autonomy or consent.

\citet{erscoi_pygmalion_2023} discuss how the concept of
\textit{Pygmalion Displacement}, derived from the myth of the Cypriot king who
preferred the marble statue Galatea over real women, remains a consistent trope
in science fiction as well as in recent technological advancements. Our study
indicates that in the context of generative AI, this fantasy is not only
perpetuated but also operationalized through the personalization of models that
allow users to generate synthetic female figures tailored to idealized and often
hypersexualized specifications. These representations emerge from a complex
interplay of platform-driven incentive structures and model-inherent biases,
each reinforcing the other. The underlying technology is trained on billions of
images, among them hundreds of thousands of women's bodies with partly
questionable origin~\citep{birhane_multimodal_2021}, and subsequently
fine-tuned with data often used without the consent of the individuals
represented.

\subsection{The Proliferation of Deepfakes}

As problematic as objectification and Pygmalion fantasies may be, they differ
fundamentally from the disruptive impact that ultra-realistic deepfakes can
exert on an individual's life. The consequences of sharing pornographic
deepfakes can, depending on the circumstances or cultural context of the victim,
range from a minor nuisance to suicide~\citep{gavrilovic_nilsson2019sextortion}
or homicide, specifically femicide~\citep{honor_killing_pakistan}. While this is
tragic on an individual level, the now readily available technology may have
detrimental effects on society as a whole. The ability to manipulate someone's
likeness, face, and/or voice is not only a severe violation of that individual's
dignity, it undermines the foundational principles of liberal democracies,
including autonomy, consent, trust-based exchange and the right to
self-ownership.

While platforms such as CivitAI are not responsible for the mere existence of
such technologies, they actively normalize and even incentivize their creation.
In the case of CivitAI, this occurs through a ``celebrity-creator leaderboard''.
This term is itself a misnomer, as it extends to individuals without a
significant public visual record, such as YouTubers and Instagrammers, many of
whom might not have the financial means to remove widespread AI-generated
non-consensual pornography from the internet. As discussed in
Section~2.4, our survey of 221,304 CivitAI assets found that 15.27\% (33,804
model adapters) were flagged as person of interest, meaning they are tailored to
replicate the likeness of real individuals.

Since the initial version of this paper was written, CivitAI has implemented a
complete ban on deepfake content involving real individuals, following an update
to its platform policies~\citep{maiberg_civitai_ban_real_people_content_2025}.
Notably, the public announcement framed the decision primarily in terms of
regulatory compliance rather than the recognition of underlying ethical
concerns. However, this policy change does not retroactively eliminate models
that were created, downloaded, and disseminated before the ban. Model adapters
released prior to the policy change may still remain stored locally on users'
machines, be shared through private channels, or be redistributed across
alternative platforms beyond CivitAI's control. Recent investigations suggest
that approximately 5,000 deepfake models were migrated to Hugging Face following
the CivitAI ban~\citep{maiberg_huggingface_5000_nonconsensual_models_2025}.
While Hugging Face maintains an Acceptable Use Policy that prohibits content
enabling harassment, sexual exploitation, or the creation of non-consensual
sexual material involving real
individuals~\citep{hugging_face_content_nodate}, enforcement necessarily relies
on reporting mechanisms and moderation at scale. In practice, this allows such
models to persist in obscured or fragmented forms, illustrating how policy
enforcement varies across platforms and how harmful artifacts can migrate rather
than disappear.

This redistribution underscores the complexity of the model-sharing ecosystem.
Even after formal prohibition on CivitAI, the affordances of open model
distribution enable pre-existing deepfake models to persist, migrate to other
platforms, and continue generating harmful content. The persistence of these
models underscores the path-dependent effects of early platform governance
choices, as artifacts released into open ecosystems cannot be fully recalled
even when platforms later implement more restrictive policies.

\subsection{Pathways to Harm Mitigation}

In line with constructivist perspectives on
technology~\citep{Hayles2025-HAYBTA-3}, we argue that technological
development, including generative AI, is not determined but contingent, and as
such, subject to complex sociotechnical dynamics. All technology is shaped by
the value systems from which it emerges, stakeholder decisions, and underlying
power structures~\citep{oconnor_gender_bias_constructed_2024}. In a recent
article, \citet{morton25inscription} builds on Bruno Latour's Actor-Network
Theory~\citep{latour1986visualization} to conceptualize \textit{inscription} as
the encoding of designers' and users' values into technology, extending the
concept to TTI models, to show how datasets, design choices, and platform
practices become embedded in these models and shape both their outputs and the
visual norms they reproduce. Extending this framework to TTI personalization
introduces additional layers of inscription into an already complex system.
Besides the initial biases inscribed into problematic foundational datasets, in
the model personalization process the tools employed in dataset creation (such as
auto-tagging systems), users' preferred topics, fine-tuning choices, and the
platform settings where models are eventually shared all function as
inscriptions. Each layer may reinforce values and biases introduced at earlier
stages, producing cumulative effects that diffuse responsibility and make
resulting harms difficult to identify, isolate, and address through single-point
technical or policy interventions.

Effective safety interventions require rigorous evaluation using diverse datasets
that capture the full spectrum of potential harms~\citep{bavalatti_systematic_2025},
as well as innovative responses from platform operators that can repurpose
existing technical infrastructure toward protective ends. Notably, the same
tools that contribute to bias inscription can be adapted for harm prevention.
For instance, \citet{haidra}, a decentralized compute provider, developed a
safety mechanism that makes use of DeepDanbooru's knowledge of problematic
concepts to block abusive image generations on their
hardware~\citep{horde-safety_2025}. This demonstrates how the auto-tagging
system discussed in Section~2.4 can be repurposed as a detection mechanism for
content moderation.

Despite such innovative solutions, some issues require more immediate and
explicit policy interventions, especially the growing threat of deepfakes.
Although Hugging Face restricts the sharing of models mimicking real
individuals' likenesses without explicit consent~\citep{hugging_face_content_nodate},
recent reporting suggests that deepfake models can nonetheless be found on the
platform, and it remains unclear to what extent these policies are proactively
enforced.

Even when model-sharing platforms implement policies, significant harm may
already have spread beyond their control. Individuals affected by digital
violence require both retroactive and preventative support.
\citet{stopncii} offers preventative tools that allow individuals to hash
non-consensual intimate imagery (NCII) so partnering online content and social
media platforms can then use these hashes to detect and block such content.
\citet{amiinporn} by the NGO digital dignity, supports victims by helping detect
and remove existing non-consensual content on sites that provide database
access. Additionally, \citet{facesafe} represents a proactive approach aimed at
preventing deepfakes and identity theft.

In addition, image perturbation methods such as Glaze~\citep{shan2025glazeprotectingartistsstyle},
Nightshade~\citep{shan2024nightshadepromptspecificpoisoningattacks},
Mist~\citep{liang2023mist}, and
MetaCloak~\citep{liu_metacloak_2024} aim to protect images from unauthorized
use in model training. However, as~\citet{zhao_protective_pertubation_survey_2024}
note, these protective methods exist in a constant arms race with model
development, as newer architectures and fine-tuning techniques continuously
evolve to overcome existing defenses. While such methods could be integrated
into social media platforms as optional upload filters, their effectiveness
ultimately depends on the willingness of platforms to prioritize user protection
over engagement, a commitment that is currently
lacking~\citep{qiwei2024reporting}.

\section*{Conclusion}

Patterns of structural discrimination \textit{inadvertently} introduced into
large-scale foundation TTI models are likely to become overshadowed by the
biased models and adapters \textit{deliberately} infused into the digital
ecosystem through open-source model-sharing platforms and their users. As shown
in this study, on model-sharing platforms like CivitAI biases, particularly
those reinforcing hypersexualization and gendered objectification, are amplified
and normalized. The rapid rise of NSFW content, the over-representation of young
female subjects, and the prioritization of sensational content reflect an
exploitative and potentially harmful dynamic.

Additionally, structural discrimination embedded in today's open-source TTI
tools and models has the potential to cause significant downstream harm as they
might become widely adopted and even integrated into future consumer
applications. Even more alarming is the normalization and incentivization of
deepfake model adapters, which fundamentally violate individual dignity,
autonomy, and the core democratic principles of self-ownership and consent.

While these general remarks emerge from our study, which focuses on a single
platform, we acknowledge that platform-specific design choices and community
norms may limit the generalizability of these observations. Our analysis does
not cover all relevant aspects, and the technological landscape is evolving
rapidly. Future research could examine prompting behavior in more detail and
extend this framework of analysis to emerging text-to-video models, where the
risks of ultra-realistic non-consensual deepfakes may intensify.

Even though our observations are necessarily partial and platform-specific, they
nonetheless indicate broader structural dynamics whose continuation or
redirection depends on how online communities, platform operators, lawmakers,
and society choose to intervene. Model-sharing and social media platforms both
have the capacity to implement safeguards that can limit the spread of abusive
practices such as deepfake creation and abusive imagery. Strengthening
responsible moderation, supporting ethical model development, introducing
technical safeguards, and collaborating with NGOs focused on preventing digital
violence, are necessary steps to adjust the current trajectory. Rather than
reacting after harm has occurred, all stakeholders have the opportunity to
innovate proactively and to collectively shape the future direction of
model-sharing and personalization ecosystems.

\begin{acks}
We thank Amanda Wasielewski, Tristan Weddigen, Maria-Teresa de Rosa Palmini,
Tania de León Yong for their helpful comments and suggestions. We are also
thankful to Jonas Schnabel and Yannick Schuchmann of digital dignity e.V.\ for
their valuable insights and comments. We are particularly grateful to Lasse
Scherffig and Thomas Hawranke for their guidance in shaping the initial proposal
for this study.
\end{acks}

\declaration{Author Contributions}{%
Laura Wagner was responsible for developing the main idea, conceptualizing the
study, conducting the exploratory analysis, visualizing the results, and writing
the manuscript. Eva Cetini\'{c} contributed to the methodological design,
assisted with data interpretation, and provided critical revisions to the
manuscript. Both authors reviewed and approved the final version.}

\declaration{Ethical Approval and Informed Consent}{%
This study did not involve human participants or identifiable personal data, and
therefore did not require ethical approval or informed consent.}

\declaration{Declaration of Conflicting Interests}{%
The authors declare no potential conflicts of interest with respect to the
research, authorship, and/or publication of this article.}

\begin{funding}
This work was supported by Swiss National Science Foundation (Ambizione Grant
216104).
\end{funding}

\section*{Code and Data Availability}
The code to reproduce the results in this paper is available at
\citet{wagner2025code}. The image metadata dataset (40,630,560 images) and
model metadata dataset (231,252 models) are available upon request at
\citet{wagner2025dataset40m} and \citet{wagner2025datasetmodels}, respectively.

\clearpage
\bibliographystyle{SageH}
\bibliography{main.bib}

\clearpage
\onecolumn   
\appendix
\renewcommand{\thefigure}{S\arabic{figure}}
\renewcommand{\thetable}{S\arabic{table}}
\setcounter{figure}{0}
\setcounter{table}{0}

\section*{Supplementary Material}

\subsection*{S1. Supplementary Figures for Section 2.1:
  Data Collection and Categorization}
\AtBeginEnvironment{table}{\scriptsize}
\AtBeginEnvironment{sidewaystable}{\scriptsize}

\begin{figure}[H]
  \centering
  \begin{subfigure}[t]{0.47\textwidth}
    \begin{tcolorbox}[]
\begin{lstlisting}[]
{
    "id": 40603651,
    "nsfw": false,
    "browsingLevel": 1,
    "createdAt": "2024-11-17T12...738Z"
    ...
    "stats": {
        "likeCount": 5,
    },
    "meta": {
        "prompt": polka-dotted squirrel, 
        ultra wide angle::3 BREAK
        "civitaiResources": [
            {
                "type": "checkpoint",
                "modelVersionId": 691639,
            },
                type: "lora"
        ]
    },
    "username": "9gc90",
    "baseModel": "Flux.1 D"  
}
\end{lstlisting}
    \end{tcolorbox}
    \caption{Exemplary image metadata (lines omitted) containing the text
      prompt, as well as parameters and models and adapters used in the image
      genesis.}
    \label{fig:image_metadata}
  \end{subfigure}
  \hfill
  \begin{subfigure}[t]{0.49\textwidth}
    \begin{tcolorbox}[]
\begin{lstlisting}[]
{
    "id": 600987,
    "name": "Naive Abstract",
    "type": "LORA",
    "poi": false,
    "nsfwLevel": 1,
    "stats": {
        "downloadCount": 136,
    },
    "creator": {
        "username": "9098f7f"
    },
    "tags": [
        "abstract"
    ],
    "modelVersions": [
        {
            "publishedAt": "2024-11-25T...Z",
            "hashes": {
            }
        }
    ]
}
                    
\end{lstlisting}
    \end{tcolorbox}
    \caption{Exemplary metadata of a model adapter (e.g.\ LoRA), containing a
      unique identifier, publishing date, person-of-interest flag etc.}
    \label{fig:model_metadata}
  \end{subfigure}
  \caption{Typical image and model metadata as obtained via REST API from
    CivitAI. Some lines have been omitted for clarity.}
  \label{fig:metadata_files}
\end{figure}

\begin{figure}[h]
  \centering
  \begin{subfigure}[t]{0.49\textwidth}
    \begin{tcolorbox}[]
\begin{lstlisting}[]
"metadata":{
   "ss_max_train_steps":"3940",
   "ss_tag_frequency":{
      "an image of a woman...":16,
      "a white face":6,
      "adding depth ...":3,
      "creating contrast ...":2,
      "adding a pop of color ...":2,
      "the background is ...":2,
      "an image of a woman ... 
        she is wearing a black corset... 
        she has a black glove ... 
        her eyes are blue and her ...
        in a ponytail and she's ... 
        the background is brick wall...":2
   },
   "ss_caption_tag_dropout_rate":"0.0"
}
\end{lstlisting}
    \end{tcolorbox}
    \caption{Example of metadata containing textual training data typical for
      image captions generated using the CLIP Interrogator captioning system.}
  \end{subfigure}
  \hfill
  \begin{subfigure}[t]{0.49\textwidth}
    \begin{tcolorbox}[]
\begin{lstlisting}[]
"metadata":{
   "ss_num_train_images":"108",
   "ss_tag_frequency":{
      "kasumi":216,
      "1girl":108,
      "solo":108,
      "bare shoulders":84,
      "japanese clothes":79,
      "blunt bangs":63,
      "breasts":62,
      "weapon":53,
      "hair ornament":51,
      "sword":51,
      "kimono":49,
      "pink kimono":45
   },
   "model_name":"EMS-263145.safetensors"
}
\end{lstlisting}
    \end{tcolorbox}
    \caption{Example of metadata with textual training data typical for
      captions generated using Danbooru-based image captioning systems.}
  \end{subfigure}
  \caption{Comparison of metadata extracted from downloaded model adapters'
    \texttt{*.safetensors} files, showcasing differences between CLIP
    Interrogator and Danbooru-based captioning systems. Addendum to
    Section~2.4.}
  \label{fig:metadata_safetensors}
\end{figure}

\clearpage

\subsection*{S2. Supplementary Figures for Section 2.2:
  Image-Level Analysis}

\begin{table}[H]
  \centering
  \scriptsize
  \renewcommand{\arraystretch}{1.1}
  \caption{Monthly statistics from the image-metadata dataset, November 2022
    to December 2024. The table reports the total number of generated images,
    the number classified as NSFW and non-NSFW, the average number of images
    generated per day, and the NSFW ratio.}
  \label{table:civiverse-dateset-table}
  \begin{tabular}{rrrrrrr}
    \toprule
    \textbf{Year} & \textbf{Month} & \textbf{Total Images} &
    \textbf{NSFW True} & \textbf{NSFW False} &
    \textbf{Avg.\ Daily Images} & \textbf{NSFW Ratio} \\
    \midrule
    2022 & 11 & 3451 & 419 & 3032 & 115.03 & 0.121 \\
    2022 & 12 & 5976 & 1529 & 4447 & 192.77 & 0.256 \\
    2023 & 1 & 26789 & 11051 & 15738 & 864.16 & 0.413 \\
    2023 & 2 & 65916 & 33907 & 32009 & 2354.14 & 0.514 \\
    2023 & 3 & 133000 & 67968 & 65032 & 4290.32 & 0.511 \\
    2023 & 4 & 164187 & 82499 & 81688 & 5472.90 & 0.502 \\
    2023 & 5 & 230673 & 115468 & 115205 & 7441.06 & 0.501 \\
    2023 & 6 & 251083 & 126848 & 124235 & 8369.43 & 0.505 \\
    2023 & 7 & 324691 & 162005 & 162686 & 10473.90 & 0.499 \\
    2023 & 8 & 334493 & 151302 & 183191 & 10790.10 & 0.452 \\
    2023 & 9 & 339205 & 176758 & 162447 & 11306.83 & 0.521 \\
    2023 & 10 & 388600 & 215200 & 173400 & 12535.48 & 0.554 \\
    2023 & 11 & 552638 & 322122 & 230516 & 18421.27 & 0.583 \\
    2023 & 12 & 661746 & 370372 & 291374 & 21346.65 & 0.560 \\
    \cmidrule(lr){1-7}
    2024 & 1 & 817261 & 478818 & 338443 & 26363.26 & 0.586 \\
    2024 & 2 & 926644 & 567173 & 359471 & 31953.24 & 0.612 \\
    2024 & 3 & 1328061 & 863795 & 464266 & 42840.68 & 0.650 \\
    2024 & 4 & 1854660 & 1352812 & 501848 & 61822.00 & 0.729 \\
    2024 & 5 & 2436790 & 1857087 & 579703 & 78606.13 & 0.762 \\
    2024 & 6 & 2968333 & 2358802 & 609531 & 98944.43 & 0.795 \\
    2024 & 7 & 3518577 & 2854555 & 664022 & 113502.48 & 0.811 \\
    2024 & 8 & 3958639 & 3110646 & 847993 & 127698.03 & 0.786 \\
    2024 & 9 & 4156712 & 3289692 & 867020 & 138557.07 & 0.791 \\
    2024 & 10 & 4559842 & 3680924 & 878918 & 147091.68 & 0.807 \\
    2024 & 11 & 5082820 & 4094030 & 988790 & 169427.33 & 0.805 \\
    2024 & 12 & 5539773 & 4405664 & 1134109 & 178702.35 & 0.795 \\
    \bottomrule
  \end{tabular}

  \vspace{0.3cm}
  \noindent\parbox{\textwidth}{\scriptsize%
    \textbf{N Images:} 40,630,560 \hspace{1em} (\textbf{NSFW:} 30,754,835).
    All images were posted publicly to CivitAI between November 2022 and
    December 2024. \textit{NSFW-true} images have Browsing Levels 2, 4, 8,
    or 16. \textit{NSFW-false} images have Browsing Level 1. The NSFW ratio
    is the proportion of NSFW-true images out of the total monthly count.}
\end{table}

\newpage

\begin{sidewaystable}[p]
  \scriptsize
  \centering
  \caption{Addendum to Section~2.2, MiVOLO Age and Gender Estimations
    for 2023--2024.}
  \label{appendix:table_mivolo_2023_2024}
  \renewcommand{\arraystretch}{1.2}
  \begin{tabular}{%
      p{1.2cm} p{1cm} p{1cm} p{1.1cm}
      p{1.1cm} p{1.1cm} p{1.1cm} p{1.1cm}
      p{1cm} p{1.8cm} p{1.8cm}}
    \toprule
    Month & Total Images & Total Persons & No Persons &
    \multicolumn{2}{c}{\female\ (\%)} &
    \multicolumn{2}{c}{\male\ (\%)} &
    \(\frac{\female}{\male}\) &
    \(\female\) Age (Mean\,±\,SD) & \(\male\) Age (Mean\,±\,SD) \\
    \cmidrule(lr){5-6}\cmidrule(lr){7-8}
    & & & & L1 & L2--16 & L1 & L2--16 & & & \\
    \midrule
    2023-01 & 26 & 48 & 9 & 38.46 & 61.54 & 75.00 & 25.00 & 3.25 & 22.26 (2.38) & 30.93 (4.46) \\
    2023-02 & 67 & 68 & 16 & 38.30 & 61.70 & 100.00 & 0.00 & 11.75 & 23.53 (4.95) & 34.60 (6.52) \\
    2023-03 & 133 & 117 & 30 & 37.36 & 62.64 & 45.45 & 54.55 & 8.27 & 23.94 (5.00) & 32.40 (7.02) \\
    2023-04 & 164 & 134 & 37 & 37.84 & 62.16 & 60.00 & 40.00 & 7.40 & 24.25 (6.90) & 33.90 (9.99) \\
    2023-05 & 229 & 242 & 52 & 35.53 & 64.47 & 70.83 & 29.17 & 6.33 & 24.40 (5.61) & 32.24 (8.88) \\
    2023-06 & 251 & 233 & 72 & 40.65 & 59.35 & 53.33 & 46.67 & 5.17 & 22.55 (4.41) & 30.52 (9.03) \\
    2023-07 & 317 & 271 & 75 & 39.30 & 60.70 & 64.10 & 35.90 & 5.15 & 22.81 (4.65) & 32.16 (9.92) \\
    2023-08 & 326 & 299 & 94 & 39.59 & 60.41 & 77.78 & 22.22 & 5.47 & 23.67 (6.60) & 37.69 (8.54) \\
    2023-09 & 316 & 287 & 58 & 30.53 & 69.47 & 74.29 & 25.71 & 6.46 & 23.73 (5.30) & 30.46 (11.91) \\
    2023-10 & 351 & 294 & 78 & 36.48 & 63.52 & 68.42 & 31.58 & 6.13 & 23.33 (6.33) & 34.04 (9.39) \\
    2023-11 & 560 & 474 & 134 & 27.79 & 72.21 & 41.67 & 58.33 & 6.12 & 23.13 (4.94) & 34.84 (12.05) \\
    2023-12 & 674 & 580 & 143 & 35.97 & 64.03 & 63.77 & 36.23 & 6.77 & 23.21 (4.88) & 33.06 (11.14) \\
    \cmidrule(lr){1-11}
    2024-01 & 752 & 659 & 155 & 31.49 & 68.51 & 52.22 & 47.78 & 5.61 & 23.32 (5.50) & 33.53 (8.22) \\
    2024-02 & 868 & 728 & 211 & 29.37 & 70.63 & 55.14 & 44.86 & 5.19 & 23.77 (5.65) & 32.55 (8.44) \\
    2024-03 & 1305 & 1094 & 358 & 24.61 & 75.39 & 46.21 & 53.79 & 6.31 & 24.11 (5.58) & 32.42 (10.63) \\
    2024-04 & 1733 & 1411 & 516 & 20.19 & 79.81 & 37.64 & 62.36 & 5.96 & 24.17 (4.96) & 32.06 (9.42) \\
    2024-05 & 2304 & 1769 & 704 & 17.11 & 82.89 & 25.57 & 74.43 & 6.54 & 24.14 (4.80) & 30.84 (9.42) \\
    2024-06 & 2857 & 2049 & 981 & 13.89 & 86.11 & 23.93 & 76.07 & 5.89 & 23.77 (4.79) & 28.88 (8.36) \\
    2024-07 & 3384 & 2416 & 1155 & 14.47 & 85.53 & 21.65 & 78.35 & 6.77 & 24.01 (4.96) & 30.18 (9.71) \\
    2024-08 & 3808 & 2814 & 1299 & 15.70 & 84.30 & 28.35 & 71.65 & 5.66 & 24.03 (4.90) & 30.53 (9.11) \\
    2024-09 & 3912 & 2840 & 1350 & 14.16 & 85.84 & 27.90 & 72.10 & 6.20 & 23.92 (4.54) & 31.51 (8.64) \\
    2024-10 & 4250 & 3224 & 1381 & 13.52 & 86.48 & 26.45 & 73.55 & 5.85 & 23.94 (4.79) & 31.19 (9.83) \\
    2024-11 & 4729 & 3458 & 1587 & 15.48 & 84.52 & 23.92 & 76.08 & 5.63 & 24.08 (5.24) & 30.57 (8.82) \\
    2024-12 & 5139 & 3564 & 1860 & 14.00 & 86.00 & 24.90 & 75.10 & 5.80 & 24.04 (4.83) & 30.72 (8.61) \\
    \bottomrule
  \end{tabular}

  \vspace{0.4cm}
  \centering
  \parbox{0.9\textwidth}{\scriptsize%
    \textbf{Addendum to Section~2.2:} A 0.1\% subsample (\(n=40{,}636\)) of
    the entire image dataset was used to estimate the demography of depicted
    subjects. MiVOLO was used to estimate perceived age and gender, from which
    we calculated the female-to-male ratio and average ages (with standard
    deviation) of female and male subjects. Note: person count may exceed image
    count as multiple persons may appear per image.

    \textbf{Disclaimer:} \(\female\) and \(\male\) refer to female-read and
    male-read classifications as determined by MiVOLO. These do not necessarily
    correspond to biological sex or gender identity.}
\end{sidewaystable}

\clearpage
\newpage

\subsection*{S3. Supplementary Figures for Section 2.3:
  Model-Level Analysis}

\begin{figure}[H]
  \centering
  \includegraphics[width=\textwidth]{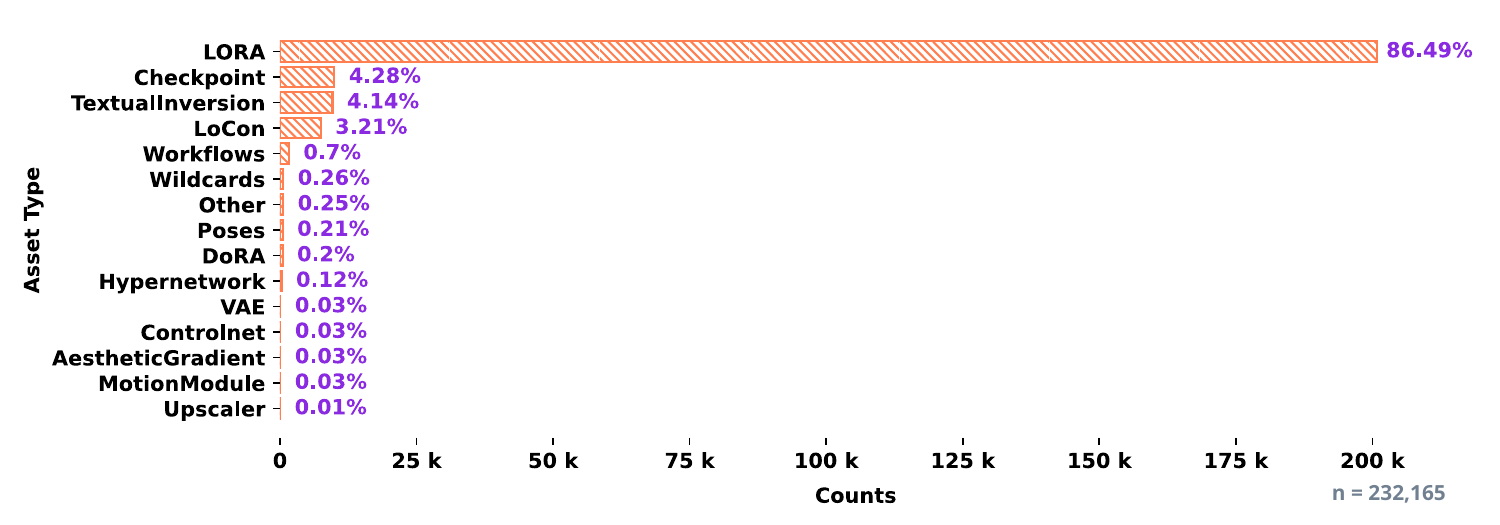}
  \caption{Distribution of asset types shared on CivitAI, excluding generated
    images. Asset types include standalone models (checkpoints), \textit{model
    adapters} (such as LoRA, LoCon, and Textual Inversion), and other resources
    such as workflows for ComfyUI, pose depth maps, or prompt ``Wildcards''.
    \textit{Wildcards} refer to placeholder tokens used to randomly construct
    prompt variations for diverse image generation. \textit{Pose depth maps}
    are structural guides used by ControlNets to constrain and compose images.}
  \label{fig:assets-shared-on-civitai}
\end{figure}

\subsection*{S4. Supplementary Figures for Section 2.4:
  System-Level Analysis}

\begin{figure}[H]
  \centering
  \includegraphics[width=0.8\textwidth]{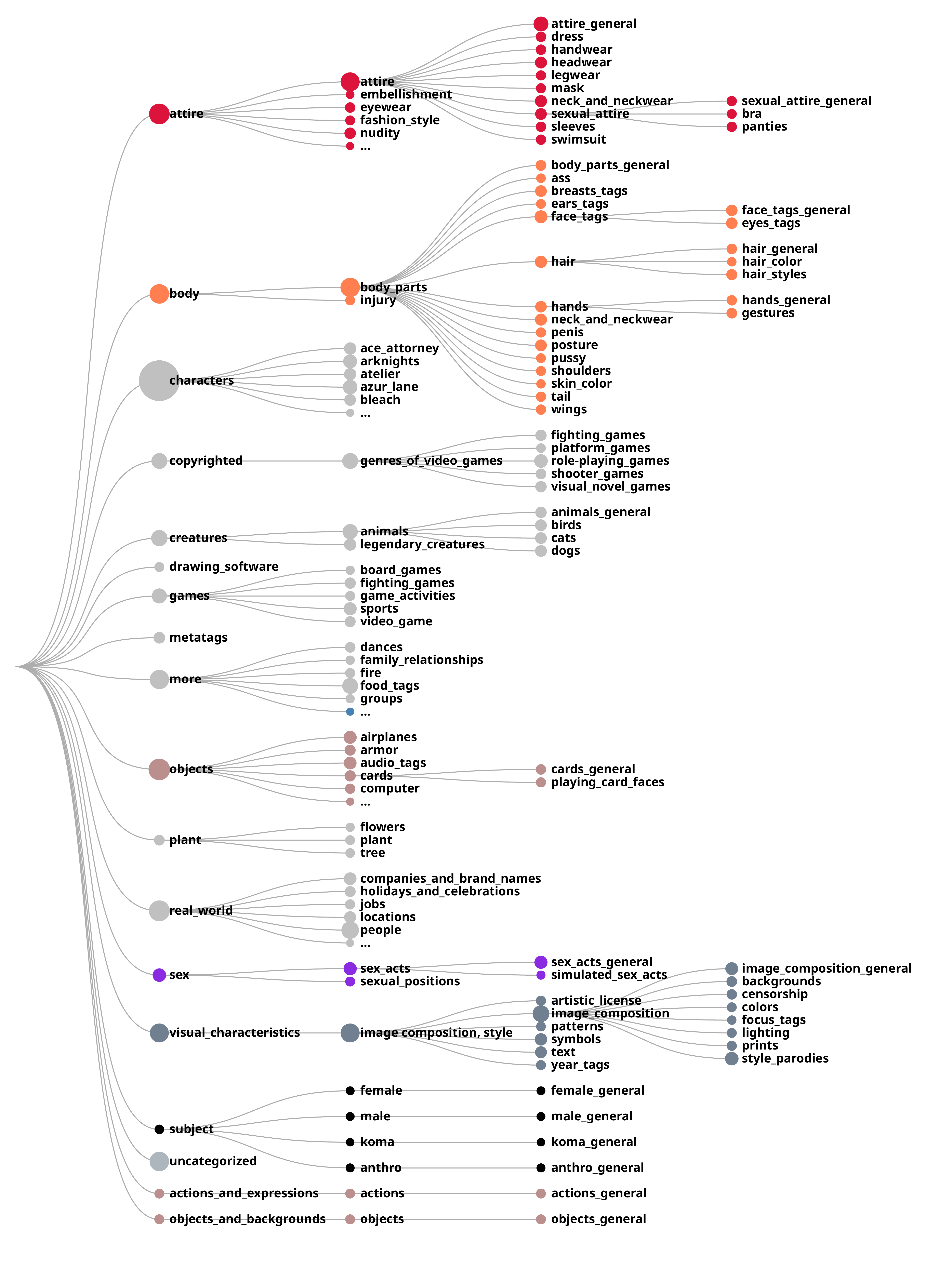}
  \caption{Danbooru taxonomy of 35,559 unique Danbooru tags as extracted from
    the tag categories and assigned the tag hierarchy from the
    \texttt{danbooru.donmai.us} tag wiki. An interactive version is available
    at: \url{https://pm-paper-b68c5e.pages.uzh.ch/Figure_13_interactive.html}}
  \label{fig:Fig_14_danbooru_taxonomy}
\end{figure}

\begin{figure}[H]
  \centering
  \includegraphics[width=\textwidth]{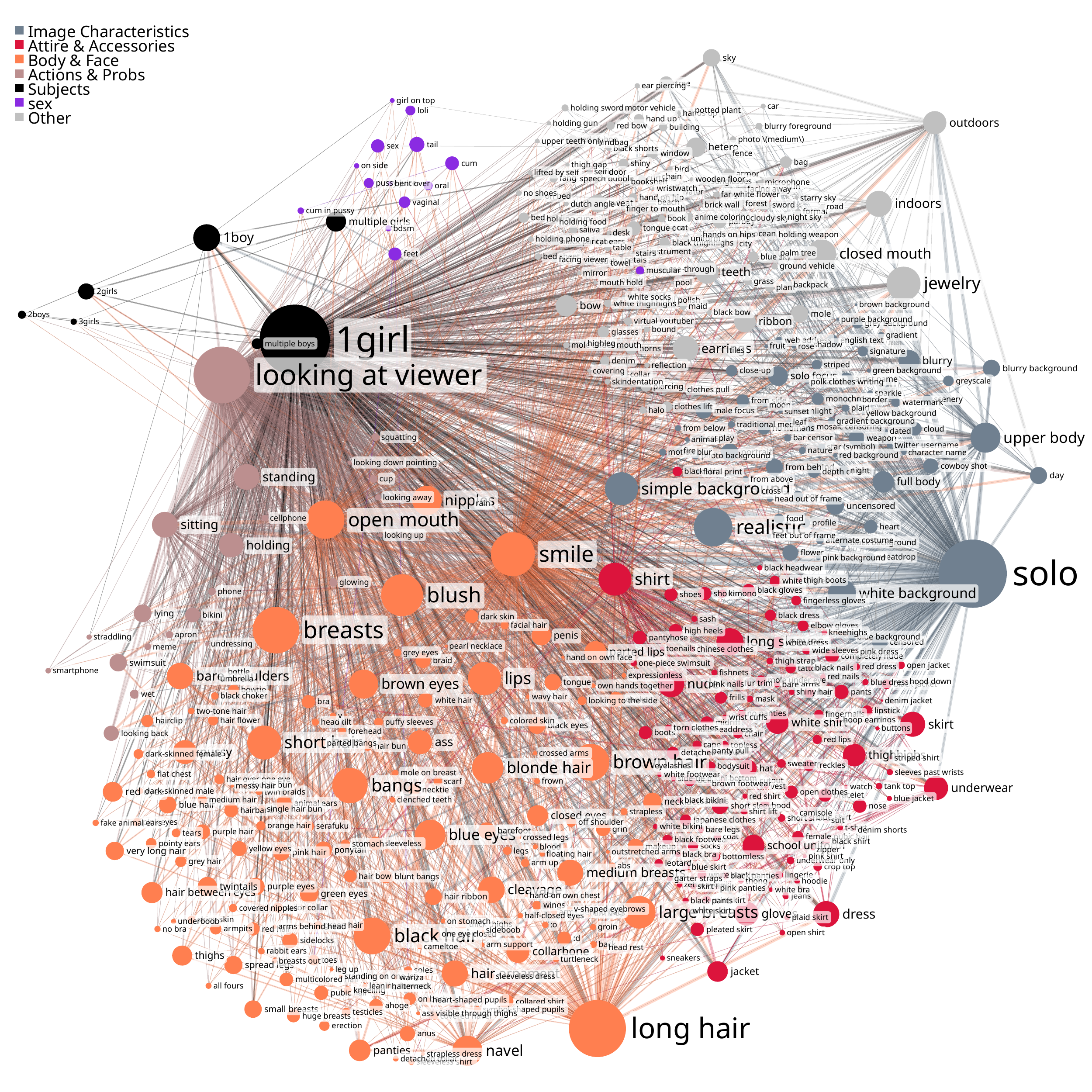}
  \caption{Co-occurrence network map of textual training data based on discrete
    tags, as described in Section~2.4. Each node represents a tag, and edges
    indicate co-occurrence relationships, with line thickness and proximity
    reflecting the strength of their association. Tags are color-coded
    according to the Danbooru taxonomy. An interactive version is available
    at: \url{https://pm-paper-viz.github.io/visualizations/figure_15.html}}
  \label{fig:Fig_15_danbooru_semantic_map}
\end{figure}

\end{document}